\documentclass[]{pasj01}
\usepackage{mathrsfs} 
\usepackage{graphicx} 
\usepackage{mediabb}
\usepackage{color}

\def\bc{\begin{center}} \def\ec{\end{center}} 
 \def\Msun{M_\odot} \def\Lsun{L_\odot}
\def\be{ \begin{equation}} \def\ee{\end{equation} }

\title{Radial Distributions of Surface Mass Density and Mass-to-Luminosity Ratio in Spiral Galaxies}  

\author{Yoshiaki \textsc{Sofue}\altaffilmark{1}  }
\altaffiltext{1}{Institute of Astronomy, The University of Tokyo, Mitaka, Tokyo 181-0015, Japan } 
\email{sofue@ioa.s.u-tokyo.ac.jp}
 
 
\date{}
  
\maketitle   
 
\begin{document} 

 \begin{abstract}  
We present radial profiles of the surface mass density (SMD) in spiral galaxies directly calculated using rotation curves (RC) on two approximations of flat-disk (SMD-F) and spherical mass distribution (SMD-S). The SMDs are combined with surface brightness (SB) using photometric data to derive radial variations of the mass-to-luminosity ratio (ML). It is found that ML has generally a central peak or a plateau, and decreases to a local minimum at $R\sim 0.1-0.2 h$, where $R$ is the radius and $h$ is the scale radius of optical disk. The ML ratio, then, increases rapidly till $\sim 0.5h$, and is followed by gradual rise till $\sim 2h$, remaining at around ML$\sim 2$ in w1 band (infrared $\lambda$ 3.4 $\mu$m) and $\sim 10\ [\Msun \Lsun^{-1}]$ in r-band ($\lambda$6200-7500 A). Beyond this radius, ML steeply increases toward the observed edges at $R\sim 5h$, attaining values as high as ML$\sim 20$ in w1 and $\sim 10^2\ [\Msun \Lsun^{-1}]$ in r-band, indicative of dominant dark matter. The general properties of the ML distributions will be useful to constrain cosmological formation models of spiral galaxies. The radial profiles of the RC, SMD, and ML are available in pdf/eps figures and machine-readable tables as an archival atlas at URL http://www.ioa.s.u-tokyo.ac.jp/$\sim$sofue/smd2018/ and as the supplementary data on PASJ home page$^\dagger$.  
\end{abstract}

\section{Introduction}  

The ultimate purpose to study rotation curves (RC) of spiral galaxies is to derive their mass distributions (e.g., Sofue and Rubin 2001; Sofue 2017). By definitiaon, RC represents the circular velocity of the disk plane, and hence treats only axisymmetric quantities as a function of the radius. Accordingly, the derived mass distribution is such integrated in the direction of rotation axis, which is the surface mass density (SMD) as projected on the galactic plane. Volume density may be estimated by dividing SMD by the vertical scale height, which needs modeling of the vertical density distribution. On the other hand, SMD has a great advantage that it can be directly compared with photometric observations, which also yields only projected surface brightness (SB) onto the galactic plane. 

The purpose of this  paper is to compile rotation curves of spiral galaxies as many as possible, and calculate SMDs using the direct method as described below. The obtained SMDs are compared with optical and infrared SB in order to obtain radial distributions of the mass-to-luminosity ratio (ML). Based on the SMD and ML for many galaxies, we discuss statistical properties about the mass distribution in galaxies. 

There have been two major methods to derive the SMD (Sofue 2017 for review). 
One is the "deconvolution method", in which observed RC is decmposed into several components such as the bulge, disk and dark halo. Each component is expressed by analytic functions of radius and density including two free parameters (mass and scale radius). Hence, we determine the six parameters of the three different types of non-linear functions by the least-squares fitting in a six dimensional parameter space. Finally the sum of thus fitted functions is vertically integrated to reach SMD. Threfore, this method requires sophysticated treatments of data and considerable computing time, so that the method is seldom to be used for SMD, despite that the method is common in studies of galactic structure. On the other hand, this method has an advantage that it is free from the edge effects as described below.

An alternative method is the "direct method", in which the SMD is directly calculated using the observed rotation curve without assuming any functional forms (Takamiya and Sofue 2000). Instead, two extreme approximations are made either that the galaxy is a flat (thin) disk or a sphere. The flat disk approximation results in under-estimation for spheroidal mass, whereas sphere over-estimates disky mass. Namely, the sphere approximation gives better estimation in the central bulge, while the outer disk is better represented by flat-disk. However, the difference of the two cases is at most a factor of $\sim 1.6$, and we consider that the true value exists in between. 
 
In sphere approximation, we must be cautious about an "edge effect" caused by the limited extent of observed RC, by which the SMD near the end radius is significantly under-estimated. This occurs due to the cylindrical (vertical) projection of spherical mass of finite radius onto the galactic plane. Similarly, although far more slightly, the flat-disk over-estimates the mass near the edge due to artificial gravitational attraction caused by cutting off of the disk mass at a finite radius. Considering that the edge effect is slighter in flat disk and that spiral galaxies are more flat-disk like than sphere, we adopt the direct method in this paper as well as in further study of the ML ratio.  

\begin{figure} 
\begin{center} 
\includegraphics[width=7cm ]{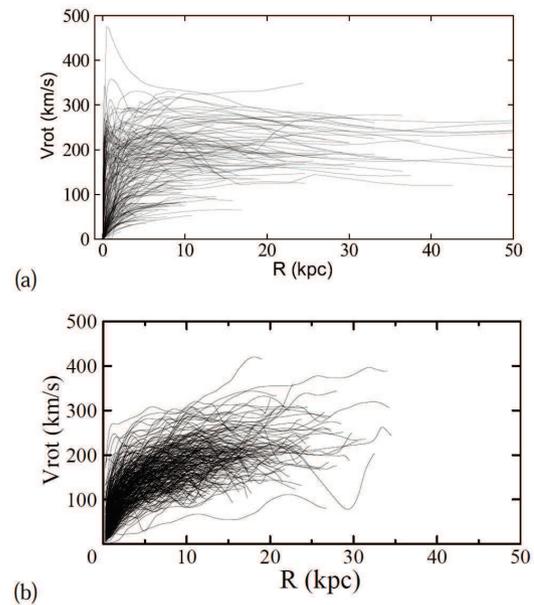}   
\ec
\caption{(a) Rotation curves from compilation by Sofue (2016), and (b) optical RCs by Courteau (1996, 1997). } 
\label{RC}    
\end{figure}

In the analyses, we use the rotation curves compiled by Sofue (2016), which we refer to as 'S-sample', and optical rotation curves obtained by Courteau (1996; 1997) referred to as 'C-sample'. 
The S-sample RCs contain those taken from 
Sofue et al. (1999;  Nearby galaxy RC atlas);  
Sofue et al. (2003; Virgo galaxy CO line survey);   
M{\'a}rquez et al. (2004; Isolated galaxy survey); 
de Blok et al. (2008; THINGS survey);  
Garrido et al. (2005; GHASP survey);  
Noordermeer et al. (2007; Early type spiral survey);  
Swaters et al. (2009: Low-SB galaxy survey); and 
Martinsson et al.(2013;  DiskMass survey). 
Several more RCs for individual galaxies were taken from the literature cited in Sofue (2016).  
All RCs used in this paper are shown in figure \ref{RC}. 

For analyses of the mass-to-luminosity ratio (ML), we employ the following archival photometric data for surface brightness of the galaxies: (a) Optical photometry in r-band ($\lambda 6800$ A (6200-7500 A)) observed by Courteau (1996, 1997), which overlaps with 'C-sample'. (b) Mid-infraraed photometry in w1-band ($\lambda$3.4 $\mu$m (2.8-3.9$\mu$m)) by Pilyugin et al. (2018), who used photometric maps from the Wide-field Infrared Survey Explorer (WISE; Wright et al. 2019), and will be referred to as 'P-sample'.

We abbreviate the quantities as SMD-F for SMD calculated by flat-disk approximation; SMD-S for SMD by spherical; ML-F for ML that is equal to SMD-F divided by surface brightness (SB); ML-S for ML equal to SMD-S divided by SB. The ML in w1-band  ($\lambda\ 3.4 \mu$m) is corrected for inclination angle of the galactic disk, and ML in r-band ($\lambda$ 6800 A) is corrected for inclination as well as the extinction. These will be abbreviated as ML-w and ML-r, respectively. The SMD, SB, and ML are measured in units of [$\Msun\ {\rm pc}^{-2}$],  [$\Lsun\ {\rm pc}^{-2}$] and [$\Msun/\Lsun$], respectively.

 Atlas of full figures and machine-readable tables of RCs, calculated SMDs, and MLs for all analyzed galaxies are available at URL$^\dagger$, http://www.ioa.s.u-tokyo.ac.jp/$\sim$sofue/smd2018/ and on the PASJ home page as supplementary data$^\dagger$.

\section{Surface Mass Density}

In this section we present SMDs calculated for rotation curves of about two hundred galaxies from the catalog of Sofue (2016), and about three hundred galaxies from the catalog of Courteau et al (1996, 1997). 
The surface mass densities, SMD-F for flat-disk approximation and SMD-S for sphere, are calculated by the following methods (Takamiya and Sofue 2000),

\subsection{Spherical approximation}

By spherical approximation, the SMD is represented by
\be
{\rm SMD_{\rm S}}=\Sigma_{\rm S}(R) = \frac{1}{2 \pi} \int\limits_R^{\infty} \frac{1}{r \sqrt{r^2-R^2}} \frac{dM(r)}{dr}dr ,
\label{eqSMDS}
\ee 
where $ M(r)=\frac{r {V(r)}^{2}}{G}$. 
The spheroidal component in the central region is well represented by this method. However, in the outer disk regions, this equation under-estimates the mass due to the edge effect near the end of the observed rotation curve. Note that SMD is a cylindrical projection of mass on the disk, but a sphere of finite radius does not cover the region away from the disk plane.    

\subsection{Flat-disk approximation} 
  
The SMD of a flat thin disk, ${\Sigma}_{\rm D}(R)$, is calculated by solving the Poisson's equation assuming a disk of negligible thickness (Freeman 1970; Binney\&Tremaine 1987), and is given by
$$
{\rm SMD_{\rm F}}={\Sigma}_{\rm F}(R) =\frac{1}{{\pi}^2 G} \times
$$
\be
\left[ \frac{1}{R} \int\limits_0^R 
{\left(\frac{dV^2}{dr} \right)}_x K \left(\frac{x}{R}\right)dx 
+ \int\limits_R^{\infty} {\left(\frac{dV^2}{dr} \right)}_x K \left
(\frac{R}{x}\right) \frac{dx}{x} \right].
\label{eqSMDF}
\end{equation}
Here, $K$ is the complete elliptic integral, which becomes very large when $x\simeq R$. Generally, this method gives a more reasonable estimates to the mass of a spiral galaxy.

\subsection{SMD plotted against radius}

\begin{figure} \bc    
\includegraphics[width=6.5cm]{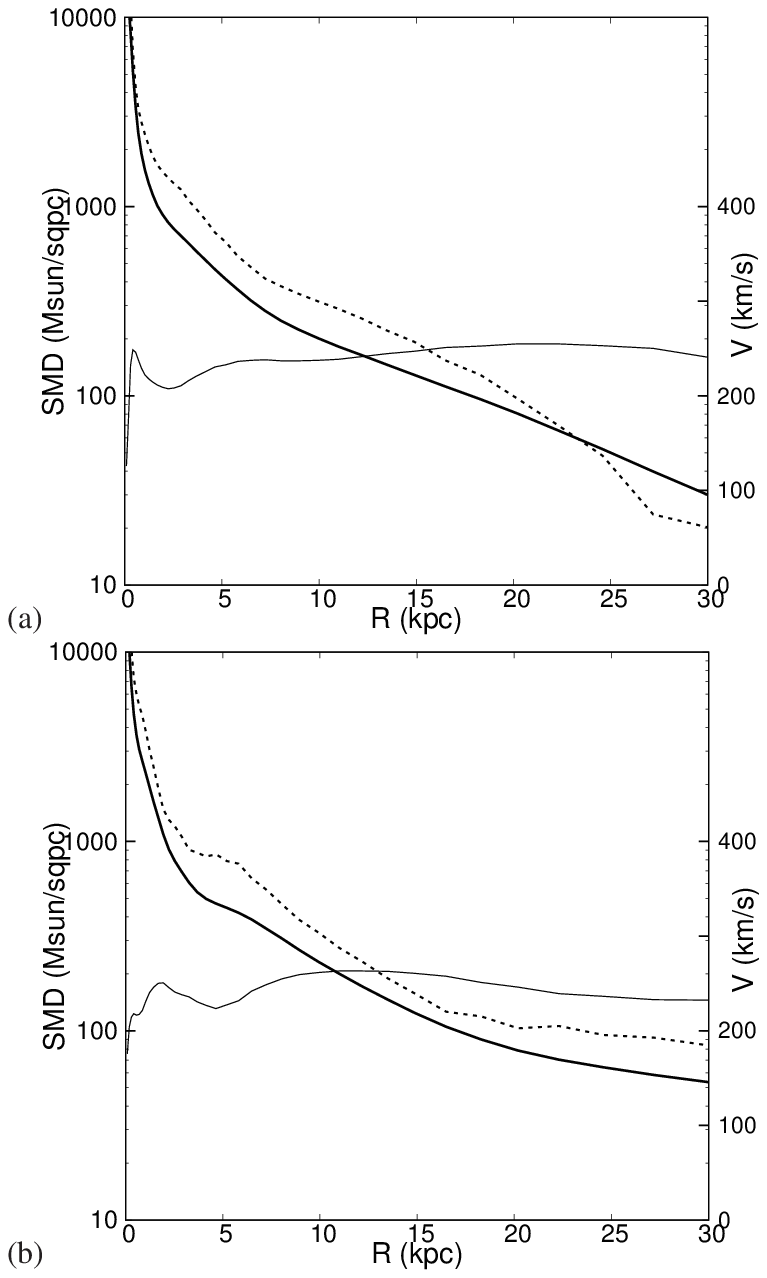}   
\ec  
\caption{ (a) Radial profiles of SMD-F (thick line), SMD-S (dashed line) and RC (thin line) for the Milky Way, and (b) M31. 
} 
\label{smd-mwm31}
%\end{figure} 

%\begin{figure} 
\bc         
\includegraphics[width=6cm]{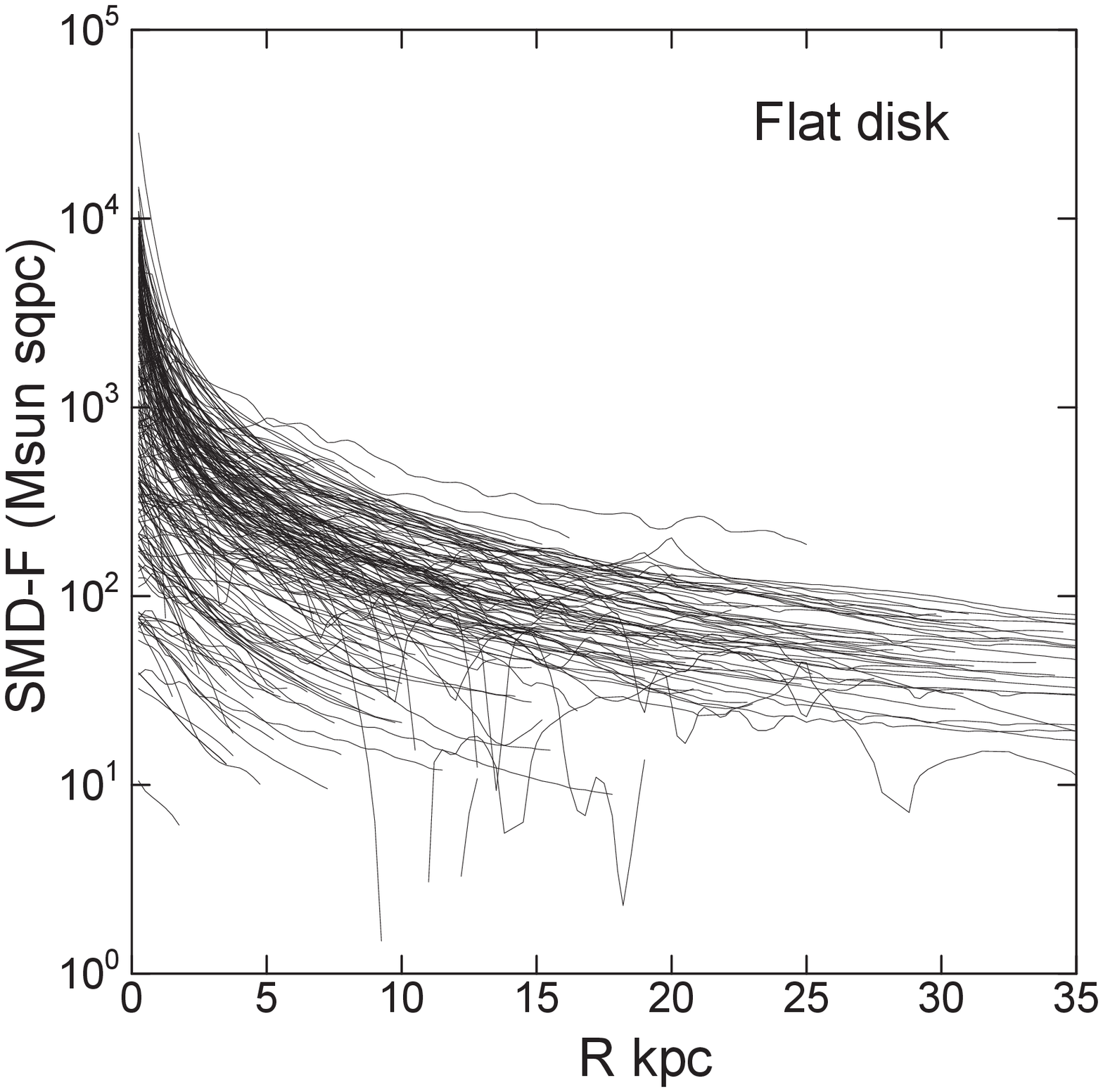}    
\includegraphics[width=6cm]{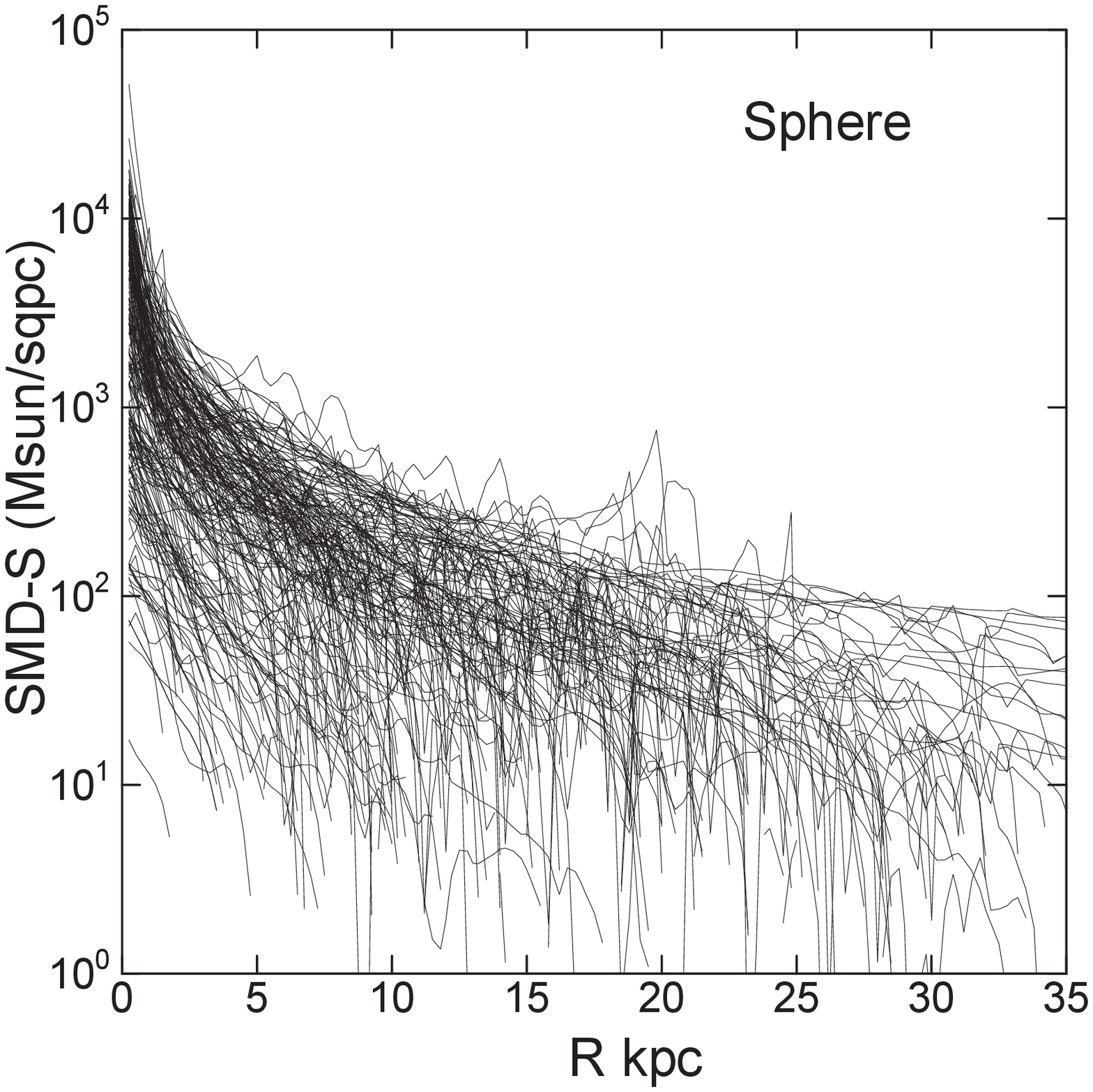}     
\ec
\caption{Surface mass densities for flat-disk (SMD-F) (top) and sphere (SMD-S) (bottom) approximations plotted against radius $R$ for S sample galaxies (Sofue 2016).} 
\label{smdall-S}    
\end{figure}

Figure \ref{smd-mwm31} shows calculated SMD-F and SMD-S for the Milky Way and M31 together with RC.  
The radial variations of SMDs are similar to each other, revealing highly concentrated central bulge, exponentially decreasing disk, and slowly decreasing outskirts representing the dark halo. 

In the Appendix by figures \ref{fig-sample-smd-S} and \ref{fig-sample-smd-C} we present samples of individual plots. Results for all galaxies are presented as an archival atlas at URL$^\dagger$.

\begin{figure*} 
\bc          
 \includegraphics[width=15cm]{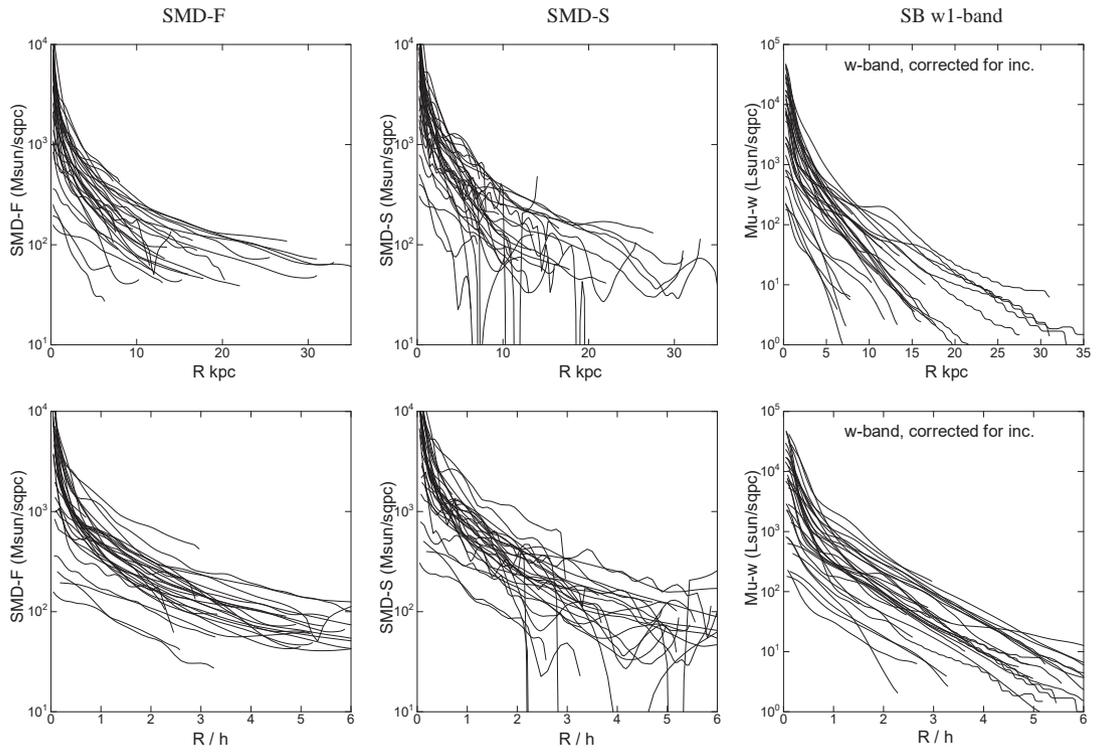}     
\ec 
\caption{[Upper panels] SMD-F, SMD-S, and SB ($=\mu$) in w1-band for S/P sample galaxies (3.4$\mu$m; Pilyugin et al. 2014) corrected for $i$ plotted against $R$, and [lower panels] same but plotted against $R/h$ (bottom).  } 
\label{smdall-P}    
\end{figure*}

\begin{figure*} 
\bc         
 \includegraphics[width=15cm]{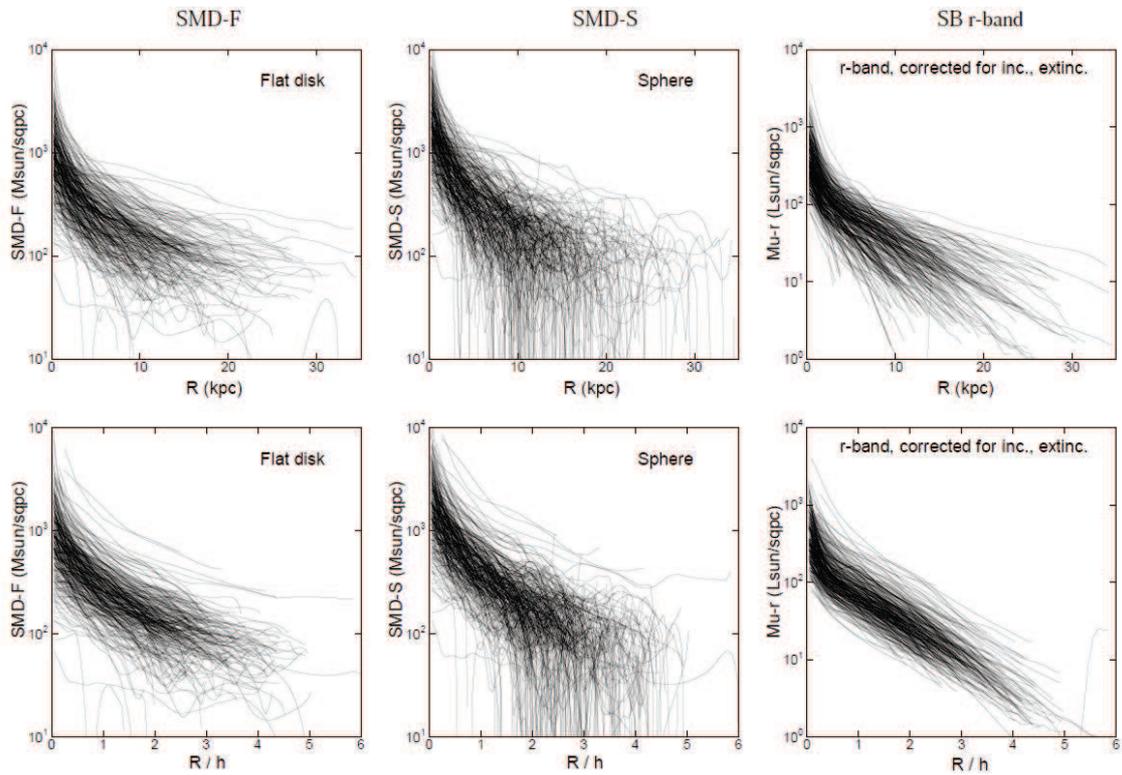} 
\ec 
\caption{[Upper panels] SMD-F, SMD-S, and SB ($=\mu$) in r-band (6200-7400A) for C-sample galaxies (Courteau 1996; 1997) corrected for $i$ and dust extinction plotted against $R$, and [lower panels] same but against $R/h$ (bottom). Note the ordered behaviors of SB in the plots against $R/h$.} 
\label{smdall-C}    
\end{figure*}

Figure \ref{smdall-S} shows SMD-F and SMD-S calculated for rotation curves of S sample galaxies (Sofue 2016). 
Figures \ref{smdall-P} and \ref{smdall-C} show SMD-F and SMD-S for P and C sample galaxies, where SMDs are plotted against $R$ as well as against the normalized radius, $R/h$, where $h$ is the scale radius given in Pilyugin et al. (2014) and Courteau (1996, 1997). Note that the plots for SMD-S show a cut-off behaviors at their end radii because of the edge effect during the integration in equation \ref{eqSMDS}, whereas SMD-F exhibits more mild behavior by equation \ref{eqSMDF}. We will use SMD-F in the following ML analyses.

In the figures, we also present radial distributions of the surface brightness (SB=$\mu$) in w (Pilyugin et al. 2014) and r-bands (Courteau 1996, 1997). Here, SBs are corrected for inclination angle of the galactic disk and dust extinction by applying the method described in the next section. 

We also plot the same against normalized radius by the scale radius, $R/h$. The distributions, particularly SB, in the outer disks show more ordered and constant gradient behavior in the plots against $R/h$.

\begin{figure}  
\bc  
\includegraphics[width=7cm]{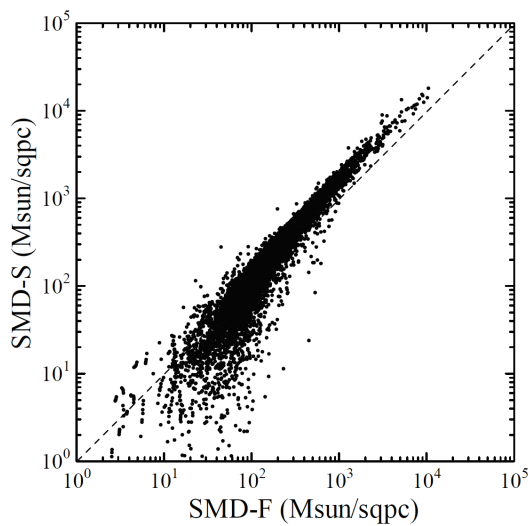}   
\ec
\caption{ 
Direct SMD-S plotted against SMD-F for all analyzed galaxies. The former generally exceeds the latter by a factor of $\sim 1.6$. Under estimation by SMD-S due to the edge-effect (see section 1) appear in the low-density regions corresponding to outer ends of rotation curves. 
}
\label{SMDFvsSMDS}      
\end{figure}

In figure \ref{SMDFvsSMDS} we plot SMD-S against SMD-F for the same galaxies, where the data have been under-sampled by a factor of 5. 
The SMD-S is systematically greater than SMD-F by a factor of $\sim \pi/2$. This is because of the spherical approximation of the mass in the SMD-S, where the mass located at large distances from the galactic plane less contributes to the radial force to balance the centrifugal force by the galactic rotation. The apparent under-estimation of SMD-S at low densities is due to the edge effect caused by the limited maximum radius of the rotation curve.

\section{Mass-to-Luminosity Ratio}

Observed w1-band (3.4 $\mu$m) SB ($=\mu$) from Pilyugin et al. (2014) in units of the solar luminosity in w1 band as [$\Lsun \ {\rm arcsec^{-2}}$] is adopted for the w1-band ML analysis. Observed data of the r-band (6200-7400A) surface brightness from Courteau (1996, 1997) in unit of [${\rm mag \ arcsec}^{-2} $] were converted to r-magnitude to luminosity by assuming that the visual absolute magnitude of the Sun is $V=4.83$ mag, and the color index between the r and v band is $V-R=0.452$ (Tanriver et al. 2016). This yields the absolute magnitude of the r-band absolute magnitude of the Sun to be $m_{\odot}( {\rm r})=4.38$ mag, which was used to calculate the surface brightness SB ($=\mu$) in r-band.
 
 \subsection{Correction for inclination and interstellar extinction}
 
\begin{figure}  
\bc  
\includegraphics[width=6.5cm]{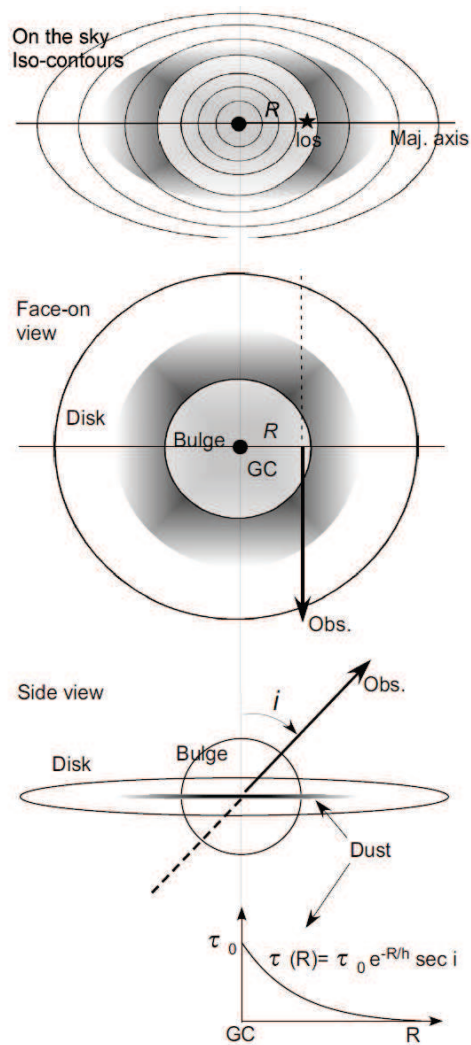} 
\ec
\caption{Correction for the inclination and dust absorption.  }
\label{illust} 
\end{figure}

 Correction for the inclination of the disk and the interstellar extinction was applied by
 \begin{equation}
{\rm SB}= \mu=\mu_{\rm obs}{ {(\cos \ i)+(1-\cos \ i)e^{-R/h}} \over { (1+e^{-\tau})/2}},
 \label{mucosi}
 \end{equation}
where, $\mu_{\rm obs}$ is the observed surface brightness at radius $R$, $h$ is the scale radius of the disk when it is fitted by an exponential function, and $i$ is the inclination angle. Figure \ref{illust} illustrates the absorption and inclination condition.

The inclination correction represented by the numerator of equation \ref{mucosi} was so applied that the central region inside $\sim h$ has a spherical brightness profile and the outer disk has an elliptical profile corresponding to the disk's inclination $i$. Thus, the outer disk brightness was corrected to a face-on value multiplying by cos $i$, whereas the central region is assumed to be spherical, dominated by the bulge brightness, yielding no or less inclination correction.

The extinction correction is approximated by the denominator of equation \ref{mucosi}. The optical depth $\tau$ along the line of sight through the inclined disk at $R$ on the major axis is approximated by
\begin{equation}
\tau =\tau_0  e^{-R/h} {\rm sec}\ i.
\label{tau}
\end{equation} 
We assume that the interstellar absorption occurs by dust layer distributed near the galactic plane, which is assumed to be sufficiently thinner than the stellar disk. Hence, light from the further side of the disk plane is absorbed, while the light from stars in front of the disk is not absorbed.
The factor $\tau_0$ in equation \ref{tau} is the vertical optical depth of the galactic disk in the galactic center.

To estimate $\tau_0$ in equation \ref{tau}, we referred to the extinction law in the solar vicinity of the Milky Way, where the r-band extinction is approximately given by (Cardelli et al. 1989; Gordon et al. 2003; Predehl and Schmitt 1995)
\begin{equation}
A_r=0.8 A_v \simeq 0.8 \ {\rm mag}\ [1.79\times 10^{21}{\rm H ~ cm}^{-2}]^{-1}
\end{equation}
.
Taking the local H density to be $0.8\ {\rm H~cm}^{-3}$ (Sofue 2017b), we obtain $A_r=1.10$ mag kpc$^{-1}$. Assuming that the thickness of the gas disk to be 100 pc, the perpendicular optical depth of the galactic disk near the Sun can be approximated by
\begin{equation}
\tau=0.1013 (n/n_0) =0.1013\ e^{-(r-R_0)/h_{\rm MW}},
\end{equation}
where $n$ and $n_0$ are the gas density in the disk at galacto-centric distance $r$ and $r=R_0=8 $ kpc, respectively, which are assumed to be expressed by an exponential function of the radius, and $h_{\rm MW}$ is the scale radius of the Galaxy. For $h=3.5$ kpc, we obtain the optical depth at the Galactic Center to be $\tau_0 (r)=0.996$. We, then, generalize this value to the galactic centers in the analyzed galaxies, and express their optical depth by equation \ref{tau}. 
The optical depth in the mid infrared w1-band is assumed to be negligible, $\tau_0(w)\simeq 0$.

These approximations may yield uncertainty in the resulting SB ($=\mu$) values by a factor of $\sim 2$, mainly due to the too much simplified extinction correction, that may be variable with radii and depend on galaxy types. Thus, we must remember that the ML ratio in this paper includes such uncertainty in addition to the uncertainty arising from the SMD-F and SMD-S selection, which has systematic uncertainty of a factor for $\sim 1.5$. 
However, these uncertainties are systematic varying from a galaxy to another, but not sensitive to the radius within a single galaxy.
Therefore, the ML ratio diagrams presented in this paper may be used to discuss the general property of the radial variation, but may not be used for analysis of their absolute values, which may be uncertain within a factor of $\sim 3$. 

\subsection{Plot of SB vs SMD}

Applying the above inclination and extinction corrections, we obtained radial distributions of SB perpendicular to the galactic planes in unit of [$\Lsun\ {\rm pc}^{-2}$]. 
The thus calculated radial profiles of SB for P and C sample galaxies are shown in figures \ref{smdall-P} and \ref{smdall-C}. 

\begin{figure} 
\bc      
\includegraphics[width=7cm]{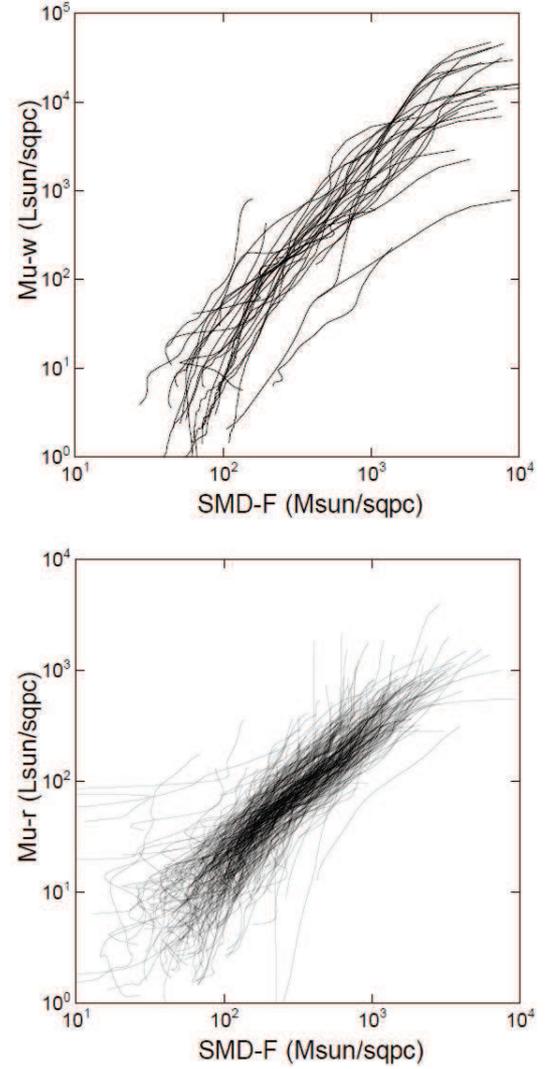}  
\ec
\caption{(a) SB ($=\mu$) vs SMD-F for S/P sample in w1-band, and (b) same for C sample in r-band.  } 
\label{SB-SMD}    
\end{figure}

Figure \ref{SB-SMD} shows the obtained SB values in w and r-bands plotted against SMD-F for P and C sample galaxies. It is trivially shown that SB increases with SMD in all the galaxies, although the correlation is not necessarily linear.

\subsection{Radial variation of ML}

\begin{figure}  \bc
\includegraphics[width=7cm]{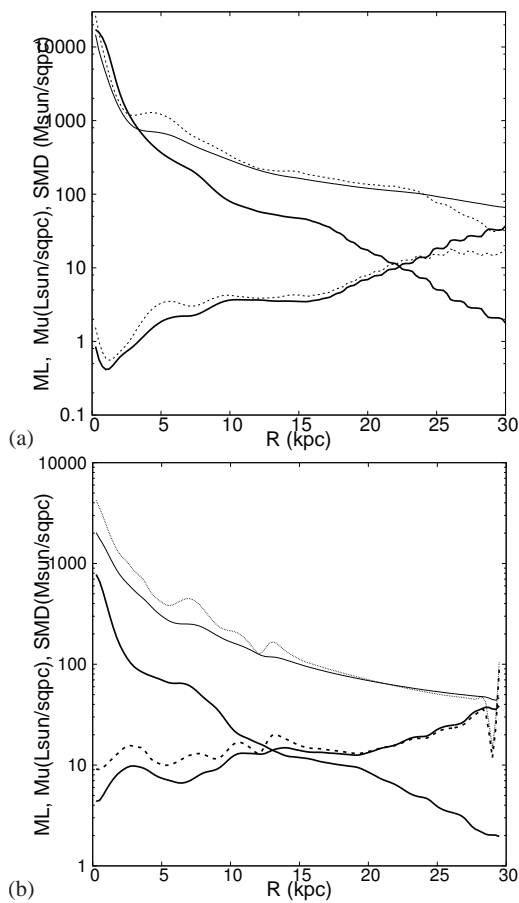}  
\ec 
\caption{(a) ML-F (thick increasing line), ML-S (thick dash), SB ($=\mu$) (thick decreasing line), SMD-F (thin decreasing line) and SMD-S (thin dash) for NGC 1097 in w1-band from S/P sample, and (b) same for UGC 14 in r-band from C sample. } 
\label{ML_U14} 
\end{figure}

The mass-to-luminosity ratio at a radius $R$ is calculated by
\be
{\rm ML(R)={{SMD(R)} \over {SB(R)}}},
\ee
where the value has the dimension of [$M_\odot/L_\odot$] in w1 or r band.
Figure \ref{ML_U14} shows radial variations of the ML ratio in NGC 1097 in w1-band from P sample and UGC 14 in r-band from C-sample galaxies. 
In figures \ref{fig-sample-ML-P} and \ref{fig-sample-ML-C} in Appendix we show more examples of individual ML-F and ML-S for P- and C-sample galaxies. All results are presented as an archive atlas at URL$^\dagger$. 

\begin{figure*} \bc      
\includegraphics[width=15cm]{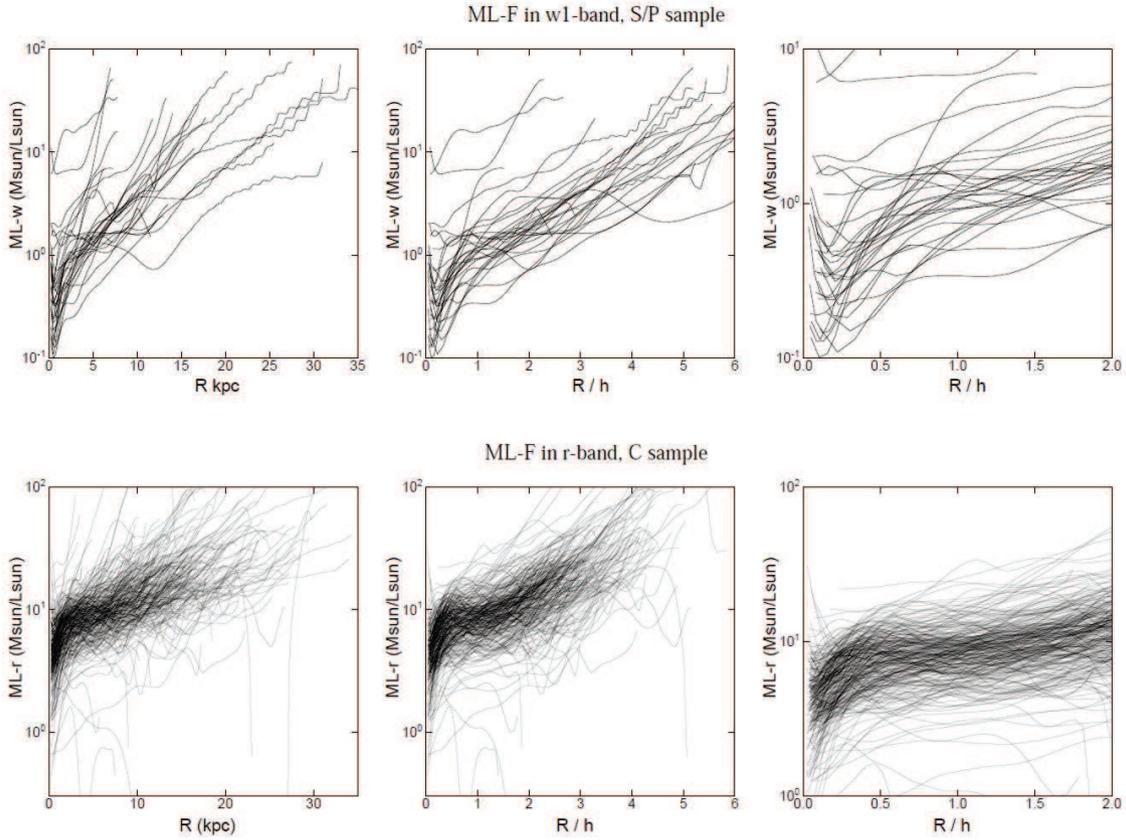}   
\ec \caption{ML-F against $R$ (left panels) and $R/h$ (middle) for P sample (w1-band) (upper panels) and C sample (r-band) galaxies (lower), and the same but enlarged for the inner region at $R/h \le 2$ (right panels). 
} 
\label{mlall}    
\end{figure*}

In figure \ref{mlall} we plot all obtained ML profiles in w1 and r-bands plotted against radius $R$ and $R/h$. In the right panels we enlarge the inner region at $R/h \le 2$. In the plots against $R/h$, the radial distributions of ML are similar to each other among the plotted galaxies. The middle panels show that the global ML increases monotonically from the disk toward the halo. 

However, as the right panels indicate, ML also increases toward the centers, attaining a central peak. Although the resolutions are not sufficient to detect and discuss the central active regions and massive objects in detail, it seems that the central regions at $R/h\le \sim 0.1 $ are dominated by high concentration of mass over the luminosity peaks. 

Leaving the center, ML decreases to local minimum at $R/h \sim 0.1$. It then increases rapidly till $R/h\sim 0.5$. Beyond this radius ML remains nearly constant at ML$\sim 2$ in w1 and $\sim 10$ in r-band, and gradually increases within the bright optical disk until $R/h\sim 2$. Beyond this radius, ML starts to increase rapidly, and reaches values as large as several tens in w1 to a hundred in r-band in the outermost regions, indicating that the halo is dominated by dark matter.

\begin{figure*} 
\bc      
 \includegraphics[width=15cm]{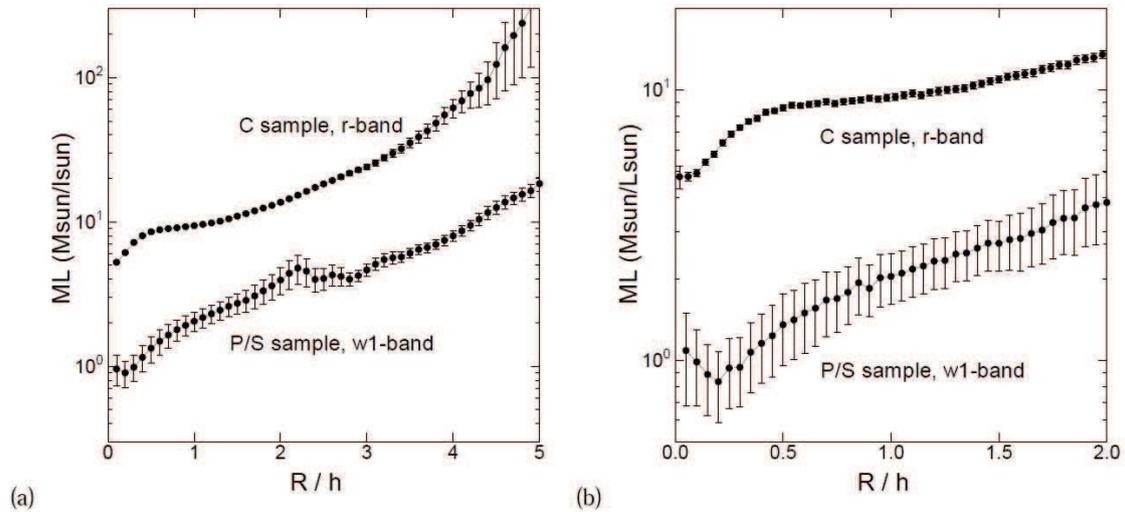}  
\ec 
\caption{(a) Gaussian-running averaged ML-F in each bin of $d(R/h)=0.1$ plotted against $R/h$ for P/S and C sample galaxies. Bars are standard errors of the mean in each bin. Smaller errors for C sample are due to larger number of data points. (b) Same, but enlarge the central region with samller bins of $d(R/h)=0.02$ for r-band, and 0.05 for w1-band.
} 
\label{MLavh}    
\end{figure*}

Figure \ref{MLavh} shows averged ML-F of the C and S/P sample galaxies after Gaussian averaging in each bin of $R/h$ with the interval of $d(R/h)=0.1$. Bars are standard errors of the mean value in each bin, which are smaller for C samples because of the larger number of data points in each bin. Note that the standard deviations are much larger as trivially seen in the original plots in figure \ref{mlall}. The general characteristics of the ML profiles as described above are equally recognized in these averaged plots, while local structures are much smoothed.
  
\subsection{ML vs SMD}

\begin{figure*} \bc      
\includegraphics[width=15cm]{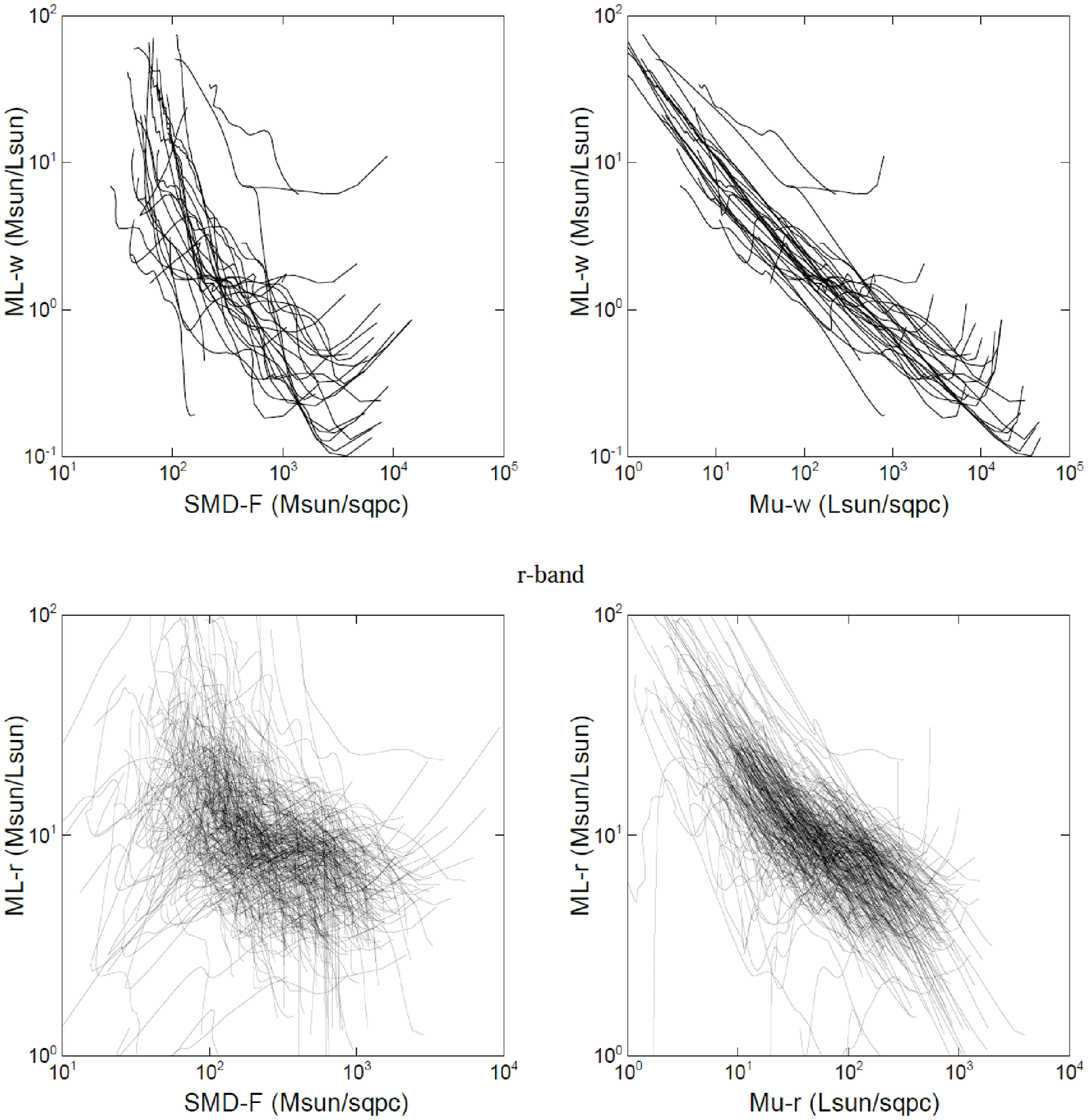}   
\ec \caption{ML-F in w1-band and r-band against SMD-F and SB ($=\mu$).   } 
\label{MLvsSMD}    
\end{figure*}

Figure \ref{MLvsSMD} shows plots of ML against SMD, and ML against SB. The ML generally decreases both with the SMD and SB. These inverse correlations manifest not only the dependence of the ML with the radius in individual galaxies, but also implies that the higher is the SB and the higher is the SMD, the smaller is the ML, regardless positions in galaxies.

\section{Discussion}

We calculated the SMD-F and SMD-S using compiled rotation curves for about five hundred galaxies. SMDs were compared with surface brightness SB ($=\mu$) in w1- and r-bands to obtain radial variations of ML ratios. The SB in r-band was corrected for inclination angle, and that in r-band was corrected both for inclination and dust extinction. The obtained results are presented as an archival atlas at URL$\dagger$. We summarize the results as follows.

\begin{itemize}
\item Galaxies show similar SMDs to each other, having a sharply concentrated central bulge, exponentially decreasing disk, and largely extended outskirt representing the massive halo.

\item SB profiles are also similar to each other. However, the outskirt of SB decreases faster than SMD, obeying the exponentially decreasing function.

\item The similarities of the radial profiles of SB and ML among galaxies are more pronounced, when they are plotted against normalized radius, $R/h$.

\item ML profiles are also similar to each other among the galaxies. It shows a central peak or a plateau, indicative of a highly concentrated mass near the nucleus. ML, then, decreases to a local minimum at $R/h\sim 0.1$ to 0.2. It, then, increases steeply till $R/h\sim 0.5$ and is followed by gradual rise till $R\sim 2h$ remaining around M$L\sim 2$ in w1 and $\sim 10$ in r-band. Beyond there, ML steeply increases toward observed edge, attaining large values, ML$\sim 20$ in w1 and about a hundred in r-band at $R\sim 5h$, representing dark matter dominant halos. 

\item ML ratios in w1-band have generally smaller values than those in r-band, manifesting larger SBs in w1-band than in r-band. 
\end{itemize}

We stress that the here obtained general characteristics of the radial distributions of the SMD and ML ratio will give a strong constraint on the formation and evolutionary scenarios of galaxies in the expanding universe  (e.g., Behroozi et al.2013; Miller et al. 2014). Particularly, the radial variation of ML will be useful to constrain the evolution models of individual galaxies, giving stronger condition than the global relation among the decomposed masses of bulge, disk and dark halos.

\vskip 5mm
{\bf Acknowledgments}:
The author thanks Prof. S. Courteau for the data base of the optical rotation curves and r-band photometry of spiral galaxies, and Prof. L. S. Pilyugin for the w1-band photometric data. The data analysis was carried out on the open use data analysis computer system at the Astronomy Data Center, ADC, of the National Astronomical Observatory of Japan (NAOJ).

\newpage
\begin{appendix}
\section{Atlas of RC, SMD and ML}

\setcounter{figure}{0} \renewcommand{\thefigure}{A.\arabic{figure}} 

\begin{figure*} 
\bc  
\includegraphics[width=15cm]{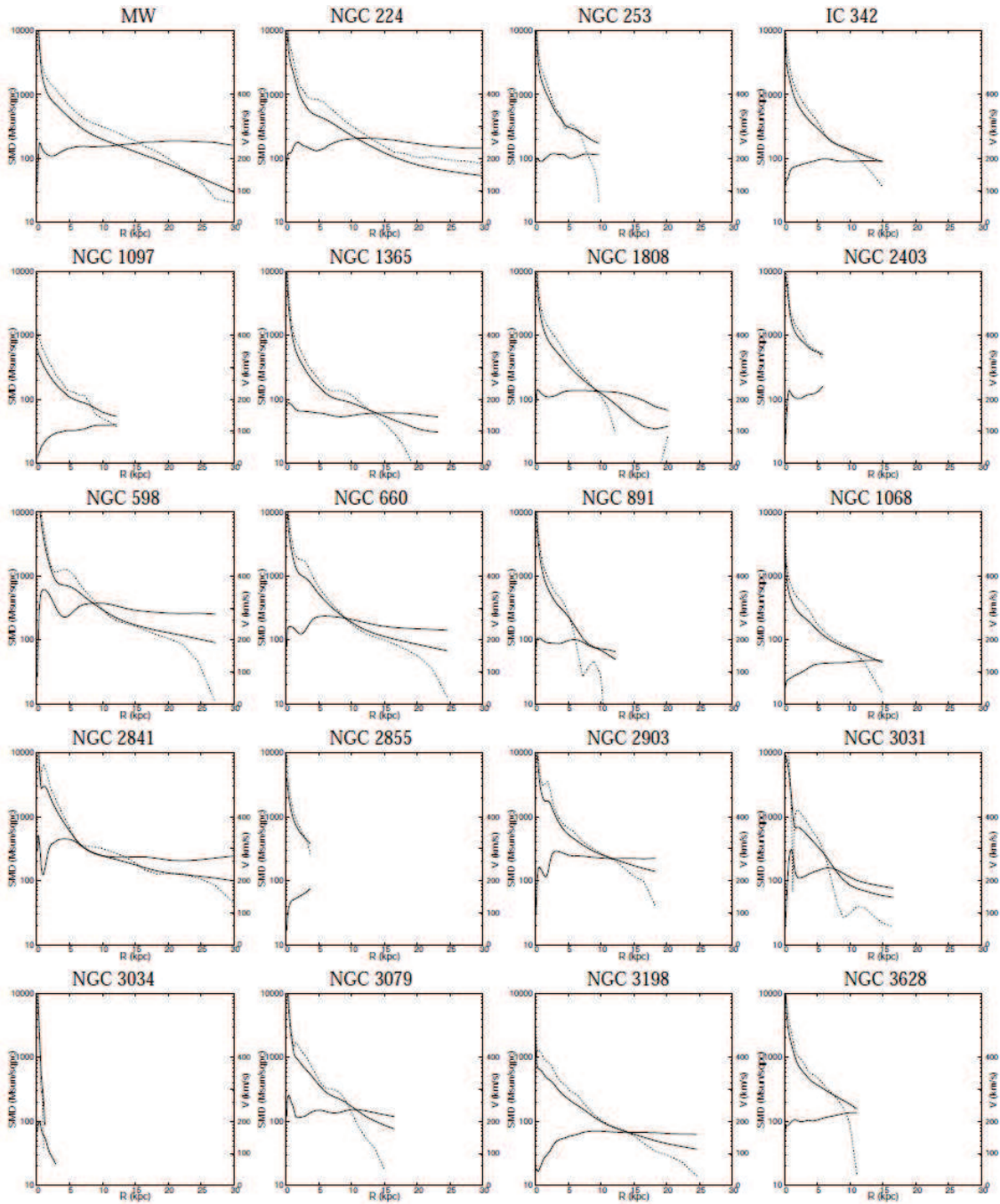} 
 
\ec 
\caption{Examples of SMD-F/S (thick/dashed lines) and RC (thin lines) from 211 S-sample galaxies (Sofue 2016). }
\label{fig-sample-smd-S} 
 \end{figure*}

\begin{figure*}  
\bc 
\includegraphics[width=15cm]{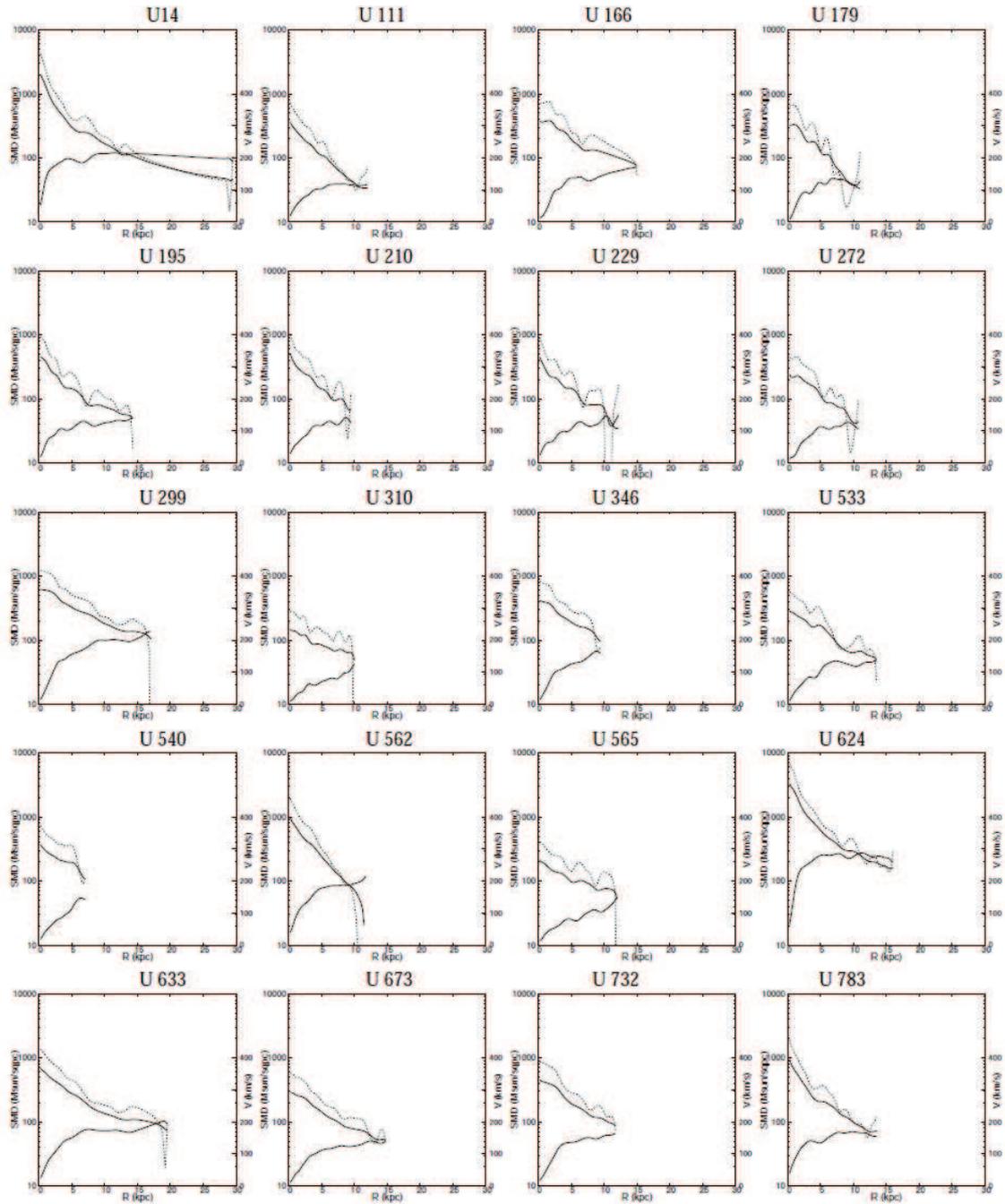}  
 \ec
\caption{Examples of SMD-F/S (thick/dashed lines) and RC (thin lines) from 289 C-sample galaxies (Courteau et al. 1996, 1997). } 
\label{fig-sample-smd-C}
\end{figure*}

\begin{figure*}  
\bc  
\includegraphics[width=15cm]{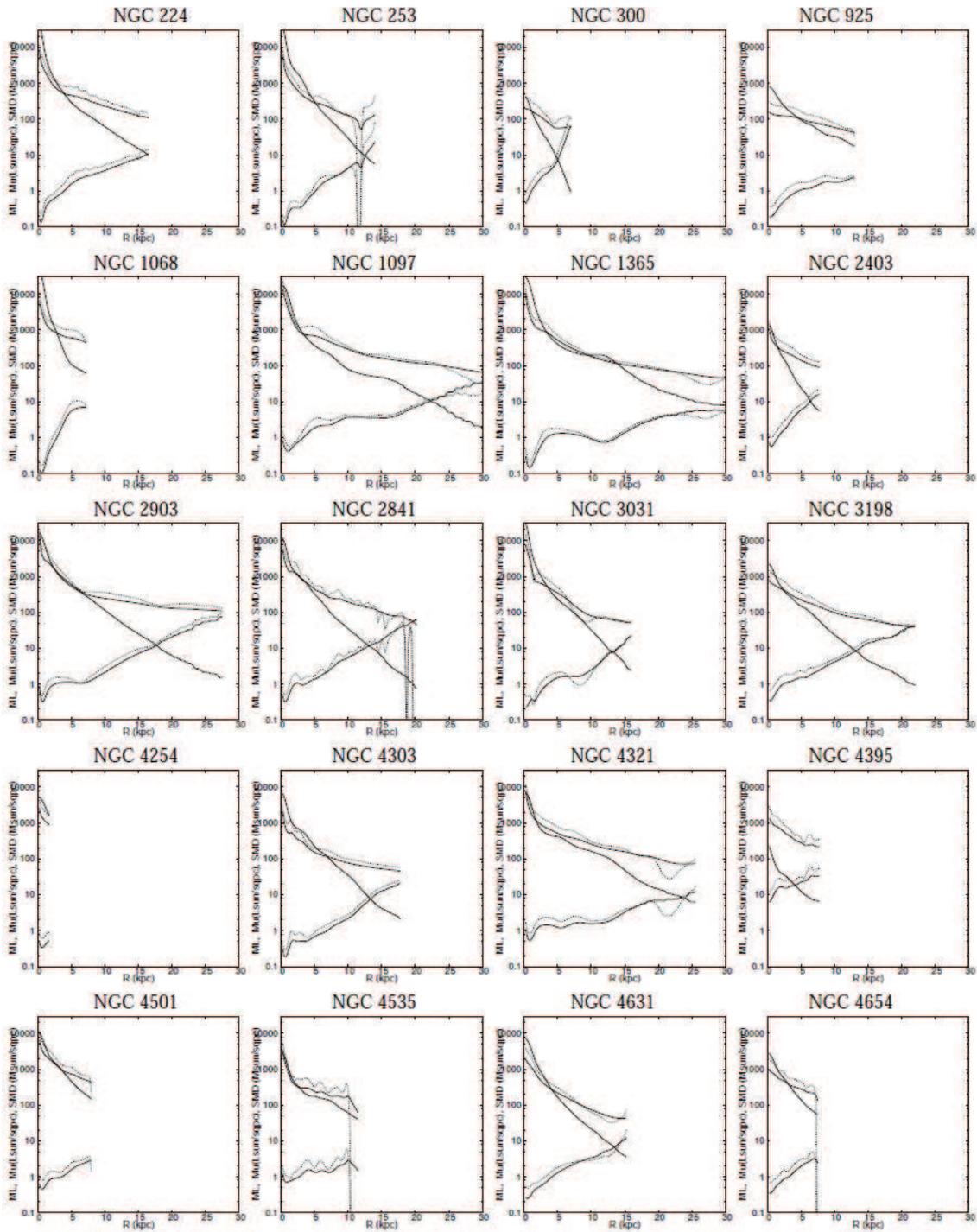}  
\ec   
\caption{Examples of w-band ML-F/S (increasing thick/dashed lines), w-band surface brightness corrected for inclination and extinction (decreasing thick lines), and SMD-F/S (decreasing thin/dash) from 31 P-sample galaxies (Pilyugin et al. 2014.   }
\label{fig-sample-ML-P} 
\end{figure*}

\begin{figure*}  
\bc 
\includegraphics[width=15cm]{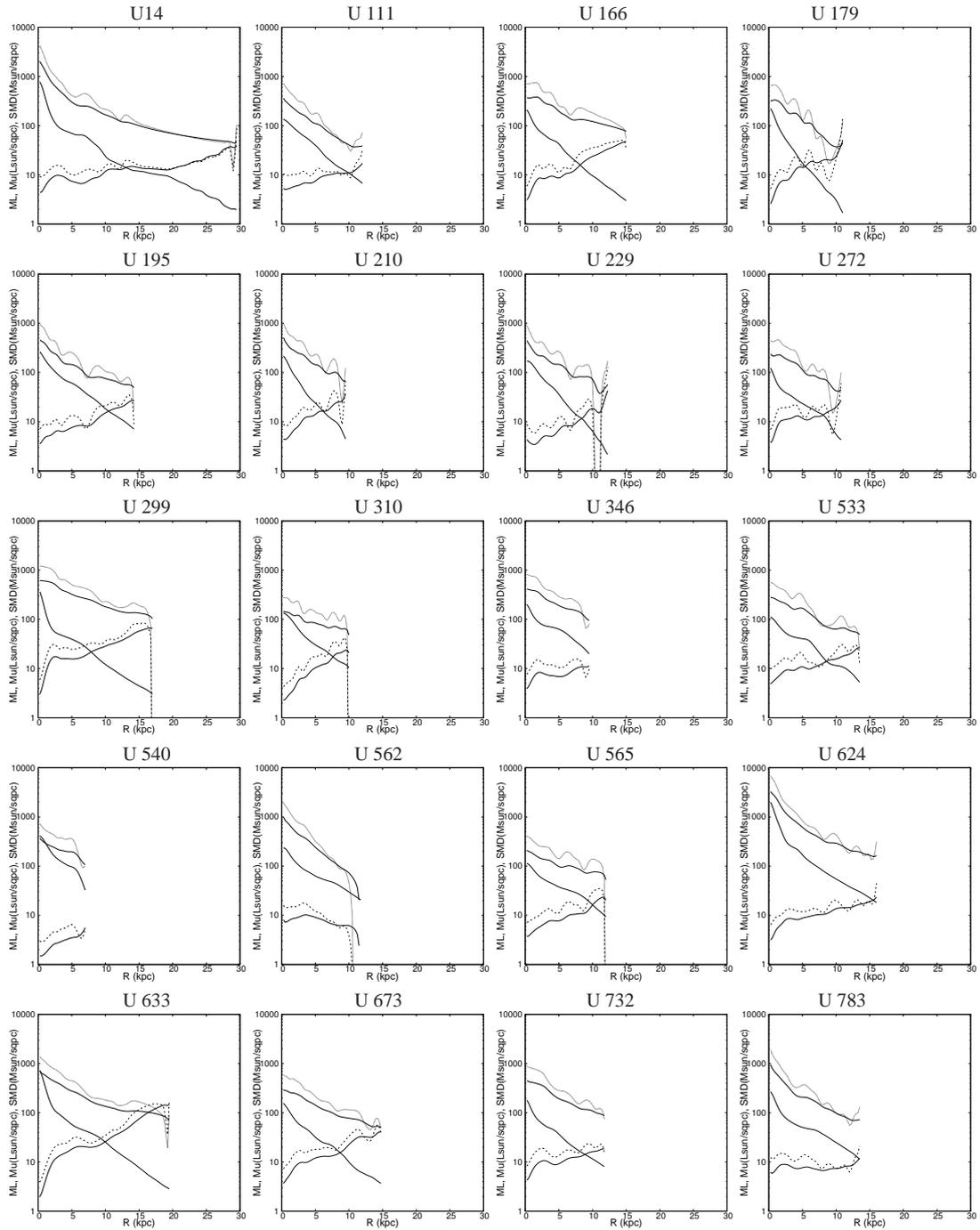} \\
 \ec      
\caption{Examples of r-band ML-F/S (increasing thick/dashed lines), r-band SB corrected for inclination and extinction (decreasing thick lines), and SMD-F/S (decreasing thin/dash) from 289 C-sample galaxies (Courteau's 1996, 1997). See URL$\dagger$ for all galaxies.}
\label{fig-sample-ML-C}
\end{figure*}

In this appendix, we present samples of the calculated SMD and ML based on the compiled RC as described in the text in figures \ref{fig-sample-smd-S}, \ref{fig-sample-smd-C}, \ref{fig-sample-ML-P}, and  \ref{fig-sample-ML-C}. Plots for all analyzed galaxies and machine-readable tables of the results are available as an archival atlas at URL$^\dagger$.

%%% SMD S-sample %%%%%%%%%%%%%%%%%%%%%%%%%%%%%%%%%%%%%%%%%%%%%%

\def\hs{\hskip 3cm}

%%%%% ML C-sample %%%%%%%%%%%%%%%%%%%%%%%%%%%%%%%%%%%%%%%%%%%%%%%%%%%%%%%%%%
 
%%% ML P-sample %%%%%%%%%%%%%%%%%%%%%%%%%%%%%%%%%%%%%%%%%%%%%%%%%%%%%%%%%%%%%%%%

%%%%% ML C-sample %%%%%%%%%%%%%%%%%%%%%%%%%%%%%%%%%%%%%%%%%%%%%%%%%%%%%%%%%%
 
\end{appendix}

\end{document}